# Measuring Vogue in American Sociology (2011-2020)

Alex Xiaoqin Yan[1], Honglin Bao[2], Tom R. Leppard[3], and Andrew P. Davis[4]
[1] Department of Sociology, Yale University
[2] Knowledge Lab & Data Science Institute, University of Chicago
[3] Data Science Academy, North Carolina State University
[4] Department of Sociology, North Carolina State University

## Abstract

This study investigates the social dynamics of knowledge production in American sociology. Departing from traditional approaches focused on citations, co-authorship, and faculty hiring, we introduce a method capturing the dynamics of networks inferred from text to explore which ideas gain traction (a.k.a *vogue*). Drawing on sociology doctoral dissertations and journal abstracts, we trace the movement of word pairs between peripheral and core semantic networks to uncover dominant themes and methodological trajectories. Our findings demonstrate that regional location and institutional prestige play critical roles in shaping the production and adoption of research trends across 114 sociology PhD-granting institutions in the United States. We show that applied research topics, such as crime and health, despite being perceived as less prestigious than theoretically oriented subjects, serve as the primary driving force behind the emergence and diffusion of trends within the discipline. This work sheds light on the institutional mechanisms that govern knowledge production, demonstrating that sociology's intellectual landscape is not dictated by simple top-down diffusion from elite institutions but is instead structured by the contextual and institutional factors that facilitate specialization and segmentation.

# Introduction

The production and diffusion of scientific knowledge are embedded within the intricate social and institutional landscapes. Sociologists argue that knowledge creation is neither neutral nor uniformly distributed but is instead embedded in systems of power and hierarchy (Uzzi and Spiro 2005, Owen-Smith and Powell 2005, Merton 1942, Bearman 1993, Burt 2004, 1987). Using semantic network analysis and natural language processing, this paper develops a novel technique to capture the diffusion and adoption of concepts and ideas in American sociology. While prior research on academic stratification has primarily relied on faculty rosters, publication records, and citation analyses (Headworth and Freese 2016, Leahey et al. 2023, Park et al. 2023, Heiberger et al. 2021), there is a need for a more fine-grained and relational illustration of how concepts emerge, circulate, and take root across institutions. Relational methodologies can capture the temporal dynamics of intellectual diffusion, allowing scholars to trace how sociological concepts rise, persist, or decline over time. These patterns align with Simmel's (1957) concept of fashion cycles, in which trends function as both mechanisms of distinction and imitation. Elite groups introduce new styles to signal status, while lower-status groups adopt these styles in an attempt to emulate them. However, once widely adopted, these styles lose their exclusivity, prompting elites to seek new forms of differentiation. This cyclical process, shaped by the interplay between class hierarchy and social status, extends beyond aesthetic trends to structure intellectual and cultural hierarchies as well.

Following this Simmelian framework, in the case of knowledge production, one might expect lower-status institutions to simply mimic the research priorities and intellectual trends set by elite departments, reinforcing a hierarchical model of knowledge production. Other reasons for elite-driven diffusion exist also - Gondal (2018), for example, finds that faculty hiring in sociology is highly status-conscious, with elite departments preferring candidates from institutions that match both their prestige and research focus. In this framework, academic hiring is shaped by both status-based social closure and shared research expertise, suggesting that elite-driven diffusion may occur not only through institutional prestige but also through disciplinary specialization. Elder and Kozlowski (2025) extend this perspective by demonstrating that status hierarchies exist not only between institutions but also within the discipline itself. Specifically, they show that high-status subfields—typically male-dominated and theoretically oriented—are concentrated in elite institutions, while lower-status, feminized, and applied subfields are found in less prestigious departments.

The hierarchy model at times reckons with the idea that the production of sociological knowledge in America has long been shaped by social, political, and economic forces. Early sociologists, particularly those from the Chicago School, developed theories and methodologies in direct response to the rapid urbanization, industrialization, and demographic shifts of the late 19th and early 20th centuries. Their research on segregation, crime, and migration reflected the realities of cities like Chicago, where economic expansion and mass immigration created new social tensions (Fine 1995; Gans 1989; Hamilton 2003). By the mid-20th century, sociology's growth was closely tied to state and federal funding, which prioritized large-scale empirical studies on population trends, social mobility, and inequality—particularly in demography and survey research—aligning the discipline with government agendas (Calhoun 2008). The establishment of land-grant universities through the Morrill Act of 1862 further institutionalized this applied orientation, as these universities were designed to conduct research that directly addressed pressing social and economic issues at both local and national levels (Turner and Turner 1990). Sociologists thus acknowledge that production is not merely dictated by prestige but is influenced by diverse institutional contexts that foster distinct intellectual trajectories.

This study extends these discussions by examining the role of meso-institutional factors, including department prestige, institutional classification, and geographic location, in structuring the movement of intellectual trends across sociology departments. Rather than viewing knowledge production as a static function of institutional status or resource constraints, this research conceptualizes it as a dynamic process, shaped by both structural hierarchies and the contingent pathways through which ideas gain traction (a.k.a *Vogue*). In doing so, We first capture vogue terms using our relational method, systematically identifying emerging concepts and tracking their diffusion. We then assign and store these terms with the institutions that adopt or generate them, creating a structured mapping of intellectual trends. Finally, we investigate how institutional classification, geographical location, and prestige influence the diffusion and adoption of vogue, assessing whether knowledge production follows a hierarchical model or diverges into distinct institutional niches.

Previous research in this tradition shows how knowledge production is not static hierarchies but rather evolving ecologies where intellectual authority is continuously reshaped (Abbott 2001, Collins 1998). Elite institutions not only control the dissemination of knowledge through prestigious journals and faculty hiring but also engage in cycles of differentiation and reintegration, as competing ideas vie for legitimacy within an ever-changing disciplinary landscape. In this framework, knowledge production is shaped by both institutional inertia and disciplinary turbulence, reinforcing the structured yet contested nature of academic fields.

We find a complementary pattern in which knowledge production is not solely dictated by elite institutions but is also shaped by local and contextual factors that foster specialization and segmentation. By demonstrating that vogue terms circulate primarily within core institutions while peripheral institutions cultivate distinct research niches, we highlight the relational nature of knowledge production: rather than a unidirectional flow from elite institutions to lower-status ones, our analysis reveals a more complex infrastructure in which knowledge is generated, contested, and localized within specific institutional ecosystems.

# Knowledge Production in Sociology

Sociologists have focused on the socially interactive nature of scientific knowledge production, particularly how new ideas emerge and diffuse through interactions with social and institutional factors (Ben-David and Collins 1996; Uzzi and Spiro 2005; Owen-Smith and Powell 2005; Leahey et al., 2023). One major stream of research examines how macro-structural forces—such as social institutions, norms, and historical processes—shape scientific knowledge production. Central to this theme is the understanding that science is not an individual pursuit but one that is socially regulated and historically embedded. Merton's (1973) analysis of the normative structure of science highlights the role of communalism and organized skepticism in fostering credible knowledge production within institutional frameworks. Kuhn's (1962) paradigm theory builds on this perspective, illustrating how scientific inquiry operates within shared frameworks that structure "normal science" and transform during periods of revolutionary change.

Collins (1998) extends this perspective, illustrating how intellectual traditions are sustained and transformed through "chains of interaction rituals," networks of scholars linked by shared ideas and interpersonal connections. Abbott (2001) complements these views by highlighting the fractal nature of disciplines, showing how fields continually reorganize and redefine their boundaries in response to broader societal forces. Scholars have demonstrated that knowledge production is dynamic, evolving through distinct life stages. According to Cheng et al. (2023), an *idea* spreads early on when it is backed by influential figures in a field and fits into collaborative research networks. These initial social dynamics provide visibility and legitimacy, allowing the idea to gain a foothold within scholarly discourse. Over time, the intellectual coherence and relevance of the idea become critical, as it is integrated into existing knowledge frameworks and consistently connected to foundational concepts. Ultimately, an idea's success is contingent on its ability to adapt to changing intellectual and social conditions, demonstrating both resonance and utility across its career. From a relational view, Kedrick et al. (2024) examine how core and periphery concepts shape the structure and relationships in knowledge production. Core concepts form the structural backbone of disciplines, maintaining stability and coherence, while peripheral concepts introduce relational diversity by connecting to new, specific ideas. Over time, as core concepts stabilize and churn decreases, scientific structures become rigid, potentially limiting innovation (Bao and Teplitskiy, 2024).

In this vein, Bourdieu (1988) conceptualizes academic fields as spaces where cultural, social, and symbolic capital are unequally distributed, reinforcing existing structures of dominance. From a socio-cultural view, Bourdieu's writing provides a compelling framework for understanding the structural and institutional sides of knowledge production. In *Homo Academicus* (1988) Bourdieu details how high-status institutions consolidate intellectual authority by dictating the criteria of legitimate scholarship. *The State Nobility* (1996) extends this view by examining how elite academic institutions function as mechanisms of social reproduction, wherein cultural and educational capital are converted into institutionalized forms of prestige. Bourdieu highlights how these institutions serve not only as sites of knowledge production but also as gatekeeping structures that regulate access to intellectual legitimacy. Through credentialing processes, professional networks, and symbolic capital, elite universities sustain their dominance by controlling what is recognized as scholarly excellence and marginalizing alternative epistemic perspectives. This process compels lower-status actors to align with dominant paradigms, reinforcing stratification within the academic field. Yet, paradoxically, the same institutions that establish these standards must also innovate to maintain their position, creating a dynamic in which conformity and distinction are mutually constitutive (Bourdieu 1984, 1988, 1996).

Simmel's (1957) theory of fashion cycles illustrates how trends function as mechanisms of both inclusion and distinction. In fashion, elite groups introduce novel styles to signal distinction, while lower-status groups adopt these styles in an effort to gain legitimacy. However, as these ideas diffuse more widely, they lose their exclusive appeal, prompting elites to seek new forms of differentiation. If academic knowledge production operates like fashion trends, we would expect similar dynamics, where intellectual trends emerge within elite institutions, diffuse downward, and either become widely adopted or are replaced by new distinctions. Applying this framework to academic knowledge production suggests two competing possibilities regarding the diffusion of intellectual trends within sociology.

One possibility is that knowledge production follows a hierarchical diffusion model, where lower-status institutions mimic the research agendas of higher-status institutions, leading to significant overlap in the concepts and ideas they prioritize. This aligns with theories of academic stratification, where institutional prestige dictates intellectual authority and access to resources (Bourdieu 1988; Merton 1973). Simmel (1957) suggests that lower-status actors adopt elite-driven styles to gain legitimacy, reinforcing intellectual convergence across hierarchies. Simmel's theory of imitation (1957) also supports this expectation, suggesting that lower-status actors adopt elite-driven styles to signal legitimacy, resulting in a convergence of intellectual trends across institutional hierarchies.

An alternative possibility is that knowledge production follows a process of distinction and differentiation, where intellectual trends remain concentrated within high-status institutions, with little diffusion to lower-status schools. Rather than simply imitating elite-driven research, lower-status institutions may develop distinct intellectual niches, leading to limited overlap in vogue terms between elite and non-elite departments. This aligns with Bourdieu's (1996) argument that elite institutions must innovate to sustain dominance, while Abbott (2001) and Kedrick et al. (2024) highlight how disciplinary structures reinforce differentiation by maintaining stability at the core while fostering innovation at the periphery. These perspectives suggest that knowledge production is shaped by institutional power and disciplinary boundaries rather than by a straightforward trickle-down model. In this sense, we state two competing hypotheses:

***H1**: If knowledge production follows a top-down hierarchical diffusion model, then lower-status schools will systematically adopt vogue terms produced by higher-status schools, leading to significant overlap in research trends. Similarly, within academic publishing, high-status journals should exhibit greater overlap with vogue terms, as they set disciplinary research agendas, while lower-status journals should mimic these trends, resulting in a similar diffusion pattern across both institutions and journals.*

***H2**: Lower-status schools develop distinct intellectual niches, driven by institutional classification, geographical proximity, and research alignment. In this model, knowledge production is shaped by local and contextual factors rather than top-down diffusion, leading to divergent research trajectories between high- and low-status institutions.*

This study builds on these theoretical foundations by empirically testing which of these two models better explains the diffusion and adoption of research trends in sociology. By leveraging semantic network analysis and natural language processing (NLP), we develop a relational and temporal approach to track how intellectual concepts emerge, circulate, and take root across institutions. Our analysis highlights how knowledge production is shaped not only by external pressures but also by internal power dynamics within the academic field. This synthesis unfolds that sociology's evolution is not merely a reflection of intellectual merit but rather a consequence of intersecting structural constraints and institutional incentives that regulate the diffusion and adoption of research trends.

## Measuring Knowledge

Bourdieu (1986) theorizes that knowledge operates as a form of symbolic capital within intellectual fields, where scholars engage in competition for recognition, status, and legitimacy. Knowledge production is not simply the generation of ideas but a process of positioning within a hierarchical structure that rewards certain types of knowledge over others. Social and institutional hierarchies govern the academic field, with scholars leveraging their cultural capital—intellectual achievements, educational credentials, and recognized expertise—to establish influence and authority. Bourdieu's concept of habitus emphasizes how these intellectual pursuits are shaped by an individual's social background and academic environment, perpetuating asymmetries of power and legitimacy in knowledge production. Knorr-Cetina (1999) introduces the concept of epistemic cultures to describe the discipline-specific norms, tools, and values that govern knowledge creation. These epistemic cultures reflect the social and technical arrangements unique to each field, showing that knowledge production is embedded in institutional and cultural contexts. Such contexts define what constitutes legitimate knowledge and shape how ideas are validated and shared, highlighting the situated nature of knowledge production. Building on this view, this study examines how research trends in sociology are constructed and diffused through relational and contextual dynamics. Expanding on these macro-structural insights, Abbott (2001) highlights how new ideas emerge at the intersections of competing subfields, where disciplinary boundaries are contested, and external influences are incorporated. This relational perspective situates disciplines as dynamic spaces, continually reshaped by cycles of fragmentation and reconstitution.

Scholars of late have been applying network analysis and text analysis to computationally model knowledge production (Yan et al., 2024). As an important component of John Mohr's formal analysis approach (1994, 1998), knowledge can be understood as a system of relations rather than isolated elements, with patterns emerging from these relations. This perspective views knowledge as fundamentally relational, where meaning and significance arise not from individual elements but from the connections and interactions among them. By applying network methods to textual data, this approach identifies how knowledge structures emerge through the relationships inferred from co-occurrences of terms in large text corpora, revealing the networked nature of intellectual frameworks (Evans and Aceves, 2016; Kozlowski et al, 2019; Macanovic 2022; Bonilowski et al, 2022). Becker et al. (2020) add to this understanding by examining how multiplex network ties—overlapping social and professional connections—facilitate the spatial diffusion of ideas. Their study of the Protestant Reformation reveals that intellectual innovations spread most effectively when supported by relational networks that bridge diverse communities and institutional hierarchies. Similarly, Cheng et al. (2023) show how the diffusion of novel scientific ideas depends on a combination of social prominence, network position, and the intellectual coherence of the ideas themselves. Over time, the influence of social factors diminishes, and ideational embeddedness within existing knowledge frameworks becomes critical for sustaining an idea's reach (Bao et al., 2024). At the micro-individual level, Foster et al. (2015) analyze how scientists navigate the tension between tradition and innovation in their research strategies. They find that while incremental contributions dominate the landscape of knowledge production, risk-taking strategies that introduce novel relationships or bridge disparate knowledge domains yield disproportionately high rewards.

Following this relational view, this paper maps the thematic structure of American sociology through semantic network analysis. We capture the career of research trends by extracting the backbone of the semantic networks and the movements of word pairs within the core-periphery structure. Our work integrates relational and contextual dynamics, showing how intellectual trends evolve over time and are shaped by institutional and geographic factors. Our work sheds light on how novel ideas emerge, gain legitimacy, and adapt within academic networks.

# Data and Method

## Data and Method Overview

The empirical data for this study is derived from a corpus of 4,747 dissertations completed between 2011 and 2020 at the 114 Sociology Ph.D.-granting institutions in the United States. Unlike the knowledge production in the physical and natural sciences, which is often characterized by laboratory experimentation, quantification, and standardization with an emphasis on reproducibility and universal applicability (Cetina 1999), the making of sociological knowledge, on the other hand, is deeply intertwined with the social and political contexts in which it is produced, interpreted, and applied (Turner 2016, Bendix 1989). This socio-cultural process highlights the passive and active nature of doing science: while the scientific community plays a critical role in validating and stabilizing facts, the broader societal context also influences which facts are pursued, accepted, or contested (Bourdieu 2001, 2010). Prior research on the production and diffusion of knowledge in sociology has relied heavily on bibliometric data, such as citation networks and journal publications (e.g., Leahey et al. 2023, Park et al. 2023, Kedrick et al. 2024). As highlighted by Heiberger et al. (2021), unlike refereed publications, which are often constrained by editorial focus and disciplinary norms, dissertations allow for greater flexibility and depth, reflecting emerging trends and niche areas of inquiry. Additionally, dissertations serve as a bridge between academic training and professional contributions, capturing the intellectual priorities that may later influence published research. Following Heiberger et al. (2021), we utilize dissertation abstracts rather than full texts to construct the corpus. Abstracts are specifically designed to provide a concise overview of the dissertation's core content, including research questions, methods, and findings. Therefore, using abstracts is both a sufficient and efficient way to extract the key information from each dissertation.

The primary approach to identifying research trends is straightforward: if a topic appears with increasing frequency in dissertations — documents that are less influenced by journal preferences, authored by graduates, and representative of American sociology — it can be considered becoming more popular. However, beyond identifying what constitutes a trending topic, it is also important to understand how these topics are studied (by looking into connected terms), as trading terms in sociology often carry multiple meanings and contexts. To achieve this, we construct semantic networks based on the dissertation corpus: A semantic network is a knowledge structure that shows how concepts (nodes) are related by a weighted edge, representing how often they are co-studied. If a connected pair moves from the periphery of the semantic network in time $T$ to the backbone of the semantic network in $T$+1, we consider the term pair to be popular. We use the first and second halves of the decade — split into five-year periods — to build two semantic networks in 2011-2015 and 2016-2020 to capture the overall trends in term usage, reflecting the shifting tastes of American sociology during the 2010s, i.e., moving from the peripheral of 2011-2015 (less likely to be co-used) to the backbone of 2016-2020 (more likely to be co-used). After collecting the set of trending term pairs, we can answer a rich range of questions on what journal more likely influences trend production, which school produced trends in the first period and adopted them in the second period, and what school-level dyadic relations shape the diffusion of trending topics from one to another. We also dig into those pairs of concepts that are always in the backbone of two semantic networks, which we term as “foundations” of American sociology, and interpret the nuance in the usage of terms between foundations and trending terms (see Appendix 1).

## Semantic Networks and the Backbone

To ensure consistency across the dataset, we applied standardized text-cleaning techniques. This process involved removing HTML tags, punctuation, and non-alphabetic characters, converting text to lowercase, and eliminating common stopwords. We customized the stopword list to retain "chapter" and "dissertation," as these terms frequently appear in abstracts but do not contribute to their substantive meaning. Finally, we applied lemmatization to reduce words to their base forms.

For text vectorization, we used Term Frequency-Inverse Document Frequency (TF-IDF), a widely adopted measure in sociological research (Macanovic 2022). TF-IDF quantifies the importance of a word within a document relative to the entire corpus by combining two metrics: Term Frequency (TF), which counts a word's occurrences in a document, and Inverse Document Frequency (IDF), which downweights words that appear frequently across the corpus. The resulting TF-IDF score increases as a word appears more often in a document but is offset by its overall frequency in the corpus, highlighting distinctive terms. We then constructed a semantic network by linking extracted terms that co-occur within the same abstract, with weighted edges reflecting co-usage frequency across the corpus. We applied backbone extraction algorithms that adaptively identify the most significant edges. Rather than imposing a fixed threshold, this method preserves edges that show statistically significant deviations from a null model in which a node's weights are uniformly distributed across its connections (for details, see Neal, 2022; Serrano & Vespignani, 2009). Extracting the backbone structure allows us to assess the relative importance of research topics and identify key drivers of their interconnections. To track the evolution of term relationships, we also conducted a close reading of original texts alongside computational analysis.

If a word pair moves from the periphery (outside the backbone) in the initial semantic network—constructed from the first five years of dissertations—to the backbone in the second, we classify it as a trending term pair in American sociology in the 2010s. This transition signals both (1) an increase in the word pair's overall usage frequency and (2) its emergence as a statistically significant structural feature within the network.

## School-level Characteristics and Dyad Relations

Given our data's dyadic nature (one school adopting vogue terms from another), we apply dyadic-cluster-robust inference (Aronow et al., 2015) with school-fixed effects for statistical analysis to well resolve the issue of potential standard error correlations within dyadic observations as well as unobserved heterogeneity at the school level. In addition to including department prestige in our model, we also add school-level institutional classifications and geo-location to examine what shared characteristics between schools facilitate the diffusion of trending terms from one institution to another. In Appendix 2, we have another set of regression analyses on what school-level characteristics make a school more likely to produce or adopt trending terms.

**Pair-Level Analysis**: The dependent variable measures the number of trending pairs flowing from one school to another. Each school is linked to 113 other schools, forming 6,441 total pairs. To examine proximity-based diffusion, we construct binary variables that capture shared characteristics between school pairs:

- Same Ranking: A categorical variable indicating whether both schools fall within the same ranking tier (e.g., Top 10, 11–20, etc.), determined by their *U.S. News* Sociology Department rankings.
- Same Size: A binary variable denoting whether both schools share the same size classification, defined as the total number of dissertations (graduates) collected from each department. Schools are categorized into three tiers—large (top third), medium (middle third), and small (bottom third). Given the highly skewed distribution of school sizes, we avoid using absolute size differences.
- Same Location: A binary variable indicating whether two schools are situated in the same geographic region, based on U.S. Census Bureau sub-regions and divisions. We opted against using state-level location measures (i.e., coding pairs as 1 if within the same state, otherwise 0) due to the fact that in many states there is only one sociology-doctoral granting institute.
- Institutional Classification: A categorical variable indicating whether a school is public, private, or land-grant.

# Results

## Research trends in the 2020s

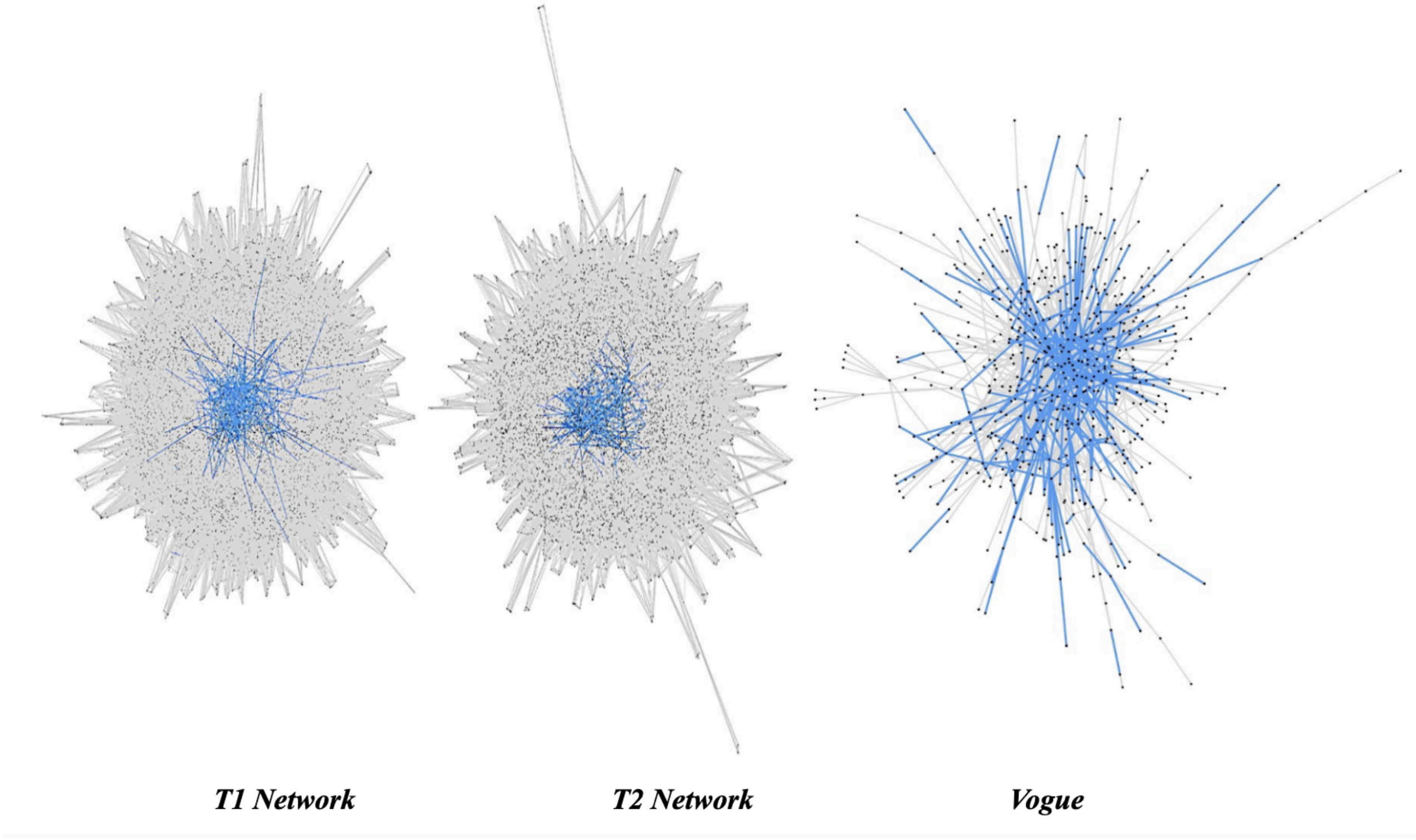

***Figure 1.*** *The global and backbone structure of dissertation semantic networks*

The emergence and development of research trends in American sociology is a dynamic process, with some ideas gaining prominence and others fading away. Figure 1 above captures the movements of research word-pairs in the 2020s through our relational methodology. Specifically, the blue areas are the backbones that extract the most significant word-pairs and the gray areas represent the word-pairs located in the periphery side of the semantic network, suggesting a less notability. The left and middle panels (T1 and T2 Networks) represent the semantic network for the first five years (2011–2015) and the second period (2016–2020). The right panel (T2 Backbone) represents the structure of research trends in the 2020s - the statistically significant, high-weight word-pairs (blue) that migrated from the periphery area in the T1 network to the backbone area in the T2 network. The grey area of the backbone of T2 is the research themes that are always on the backbone across two networks. In this sense, the research trends are inherently dynamic, driven by their relational movement across time. Likewise, their meanings are determined by the words connected to them within the semantic network rather than by static definitions. Notably 82% of vogues are arguably “weak ties”, which are non-backbone in the first five years but connected to two backbone edges.

***Table 1.*** *Representative research trends in American sociology.*

| **Representative topics** | **Connected words** |
|---|---|
| 'health' | 'objective','distress','parental','planning','incarceration', 'stigma','cognitive','reproductive','socioeconomic','young','depressive' |
| 'mental' | 'adult','physical','stress','black','psychosocial','discrimination','identity', 'distress','stigma', 'psychological','justice','adolescent','young' |
| 'gender' | 'religious','medical','violence','feminist','young','transgender','money' |

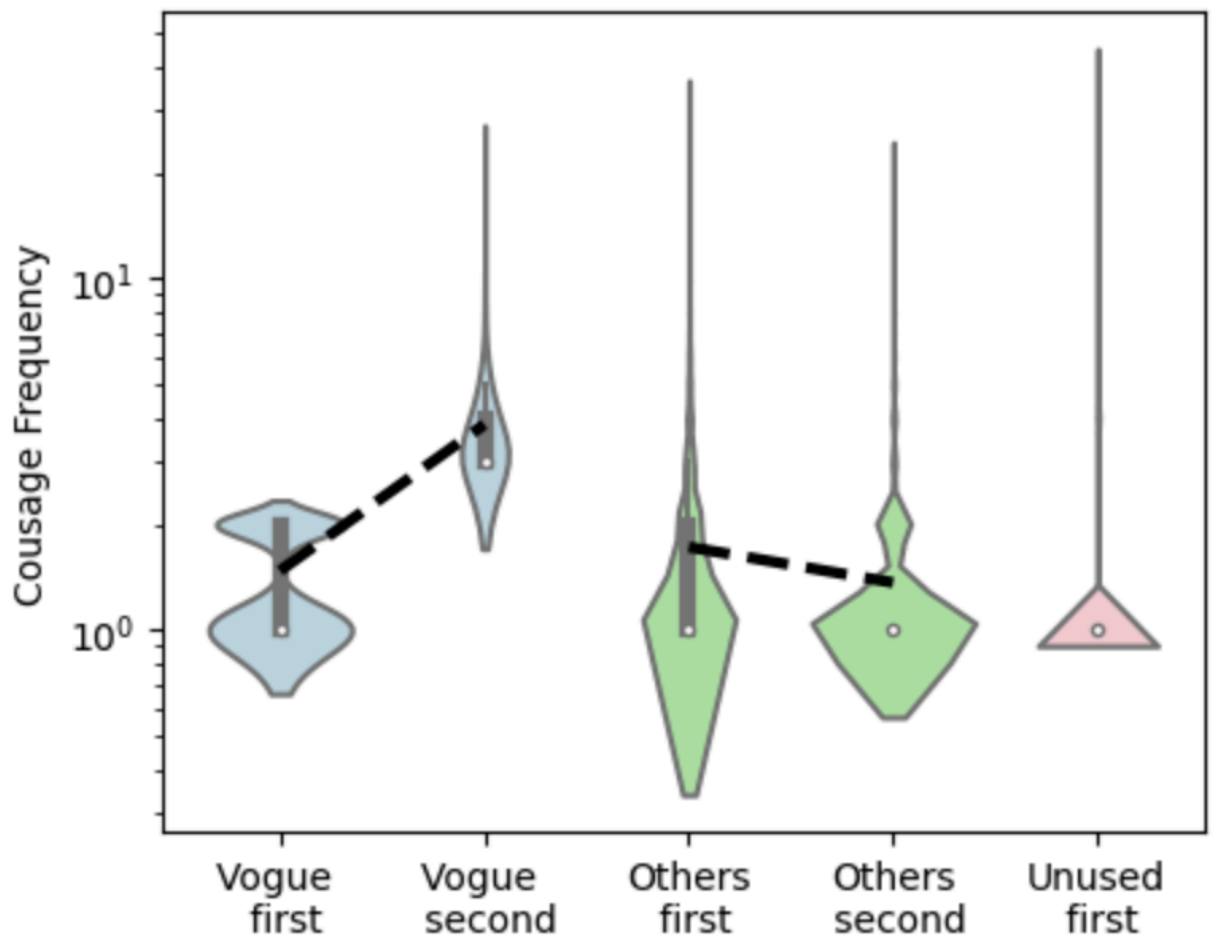

***Figure 2.*** *Vogue pairs within the semantic network and their usage dynamics.*

Figure 2 captures the fluctuations of co-usage frequency among word pairs within the semantic network, with the word pairs categorized into three groups: vogue terms, non-vogue terms, and unused terms. Vogue terms, represented by the light blue violin plots, exhibit a significant increase in co-usage frequency from the first period (T1) to the second period (T2), reflecting their growing prominence and centrality within the evolving sociology research discourse. Non-vogue terms, shown in light green, remain present in the network but experience a decline in co-usage frequency, indicating a reduced emphasis and relevance over time. Lastly, the unused terms, represented by the pink triangle, were present in T1 but disappeared entirely in T2, signifying the natural attrition of certain ideas that fail to sustain their significance.

## Journals and research trends

***Table 2.*** *Overlap between vogue and journal semantic networks.*

| Journal | Overlap | Reputation | Impact Factors |
|---|---|---|---|
| *Journal of Health and Social Behavior* | 0.005778 | 65 | 5 |
| *Demography* | 0.004012 | 81 | 4 |
| *Sociology of Education* | 0.003779 | 66 | 3.9 |
| *Journal of Marriage and Family* | 0.003681 | 71 | 6 |
| *The Sociological Quarterly* | 0.003266 | 58 | 1.2 |
| *Social Psychology Quarterly* | 0.002816 | 72 | 2.7 |
| *Gender & Society* | 0.002667 | 71 | 5.5 |
| *Sociological Inquiry* | 0.002631 | 42 | 1.9 |
| *Social Forces* | 0.002551 | 89 | 4.8 |
| *American Sociological Review* | 0.002509 | 96 | 9.1 |
| *Sociological Perspectives* | 0.002144 | 53 | 2.4 |
| *American Journal of Sociology* | 0.002143 | 94 | 4.4 |
| *Sociological Forum* | 0.002143 | 63 | 1.8 |
| *American Journal of Cultural Sociology* | 0.002042 | 37 | 2.9 |
| *Journal for the Scientific Study of Religion* | 0.002000 | 42 | 2.4 |
| *Work and Occupations* | 0.001892 | 71 | 2.9 |
| *Sociological Focus* | 0.001788 | 37 | 1.7 |
| *Social Problems* | 0.001443 | 77 | 3.2 |
| *Ethnic and Racial Studies* | 0.001435 | 53 | 2.5 |
| *Mobilization* | 0.001286 | 53 | 1.5 |
| *Sociological Spectrum* | 0.001073 | Unavailable | 1.8 |
| *European Sociological Review* | 0.000908 | 69 | 3.2 |
| *British Journal of Sociology* | 0.000897 | 70 | 2.1 |
| *International Journal of Comparative Sociology* | 0.000878 | 38 | 2 |
| *Acta Sociologica* | 0.000844 | 37 | 1.7 |
| *Social Networks* | 0.000745 | 65 | 3.1 |
| *Chinese Sociological Review* | 0.000633 | Unavailable | 4 |
| *European Journal of Sociology* | 0.000567 | Unavailable | 1.4 |
| *Symbolic Interaction* | 0.000543 | 37 | 1.6 |
| *Sociological Theory* | 0.000286 | 78 | 4.4 |
| *Sociological Methods & Research* | 0.000192 | 75 | 6.3 |
| *Sociological Methodology* | 0 | 74 | 3 |

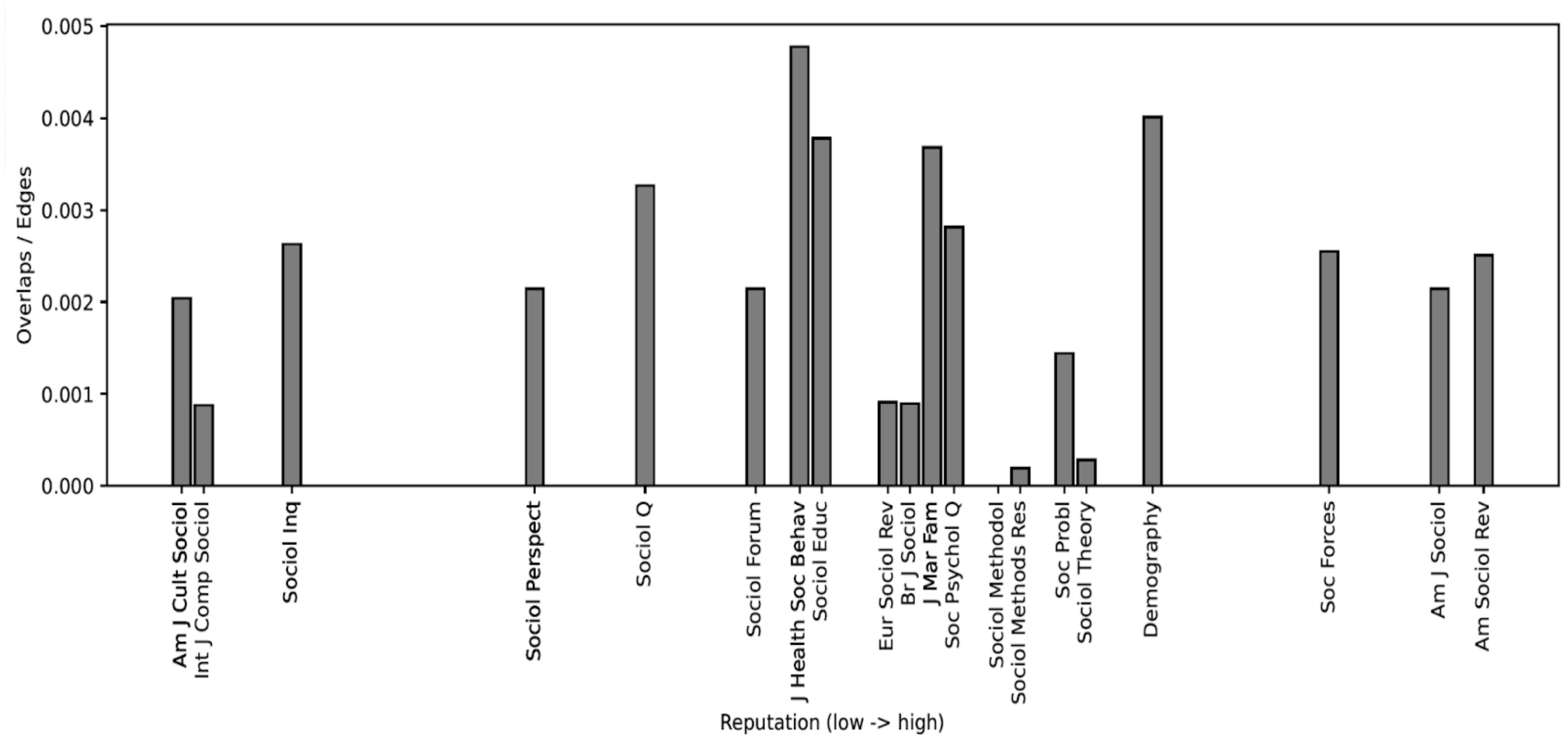

***Figure 3.*** *The most prestigious refereed journals do not produce the most fashionable research trends.*

What journals most likely impact vogue popularity? We sourced the journal reputation rankings from the website All Our Ideas[1], where sociologists were asked to evaluate and rank journals based on the question, "Where would be better for a grad student to publish a solo-authored article?" (McDonnell & Stoltz, 2020). We selected a cluster of representative journals that cover generalist, specialist, theoretical and methodological topics. Based on the selected journal list, we collected their published papers' abstracts in the first five years, built a corresponding semantic network for each journal, and used the overlap between the journal's network edges with vogue edges to approximate how journals impact vogue popularity in the second period. We normalized the overlap with vogue edges by the number of total edges of the corresponding journal's semantic network. As shown in Table 1, we found that the journals producing the most vogue are not necessarily the most prestigious journals listed in the journal reputation rankings. Rather, empirical journals with a central focus on American social issues (e.g., *Journal of Health and Social Behavior, Demography*) have the highest overlap with vogue, suggesting their critical role in research trendsetting. *American Journal of Sociology* and *American Sociological Review*, although ranked highly in their reputation score, obtained a mediocre measure in terms of producing vogue. The methodology and theory journals (*Sociological Methods & Research, Sociological Theory, and Sociological Methodology*) received the lowest score, suggesting that the knowledge infrastructure of research methods and theory has remained less central in the 2020s.

[1] http://www.allourideas.org/socpublishing

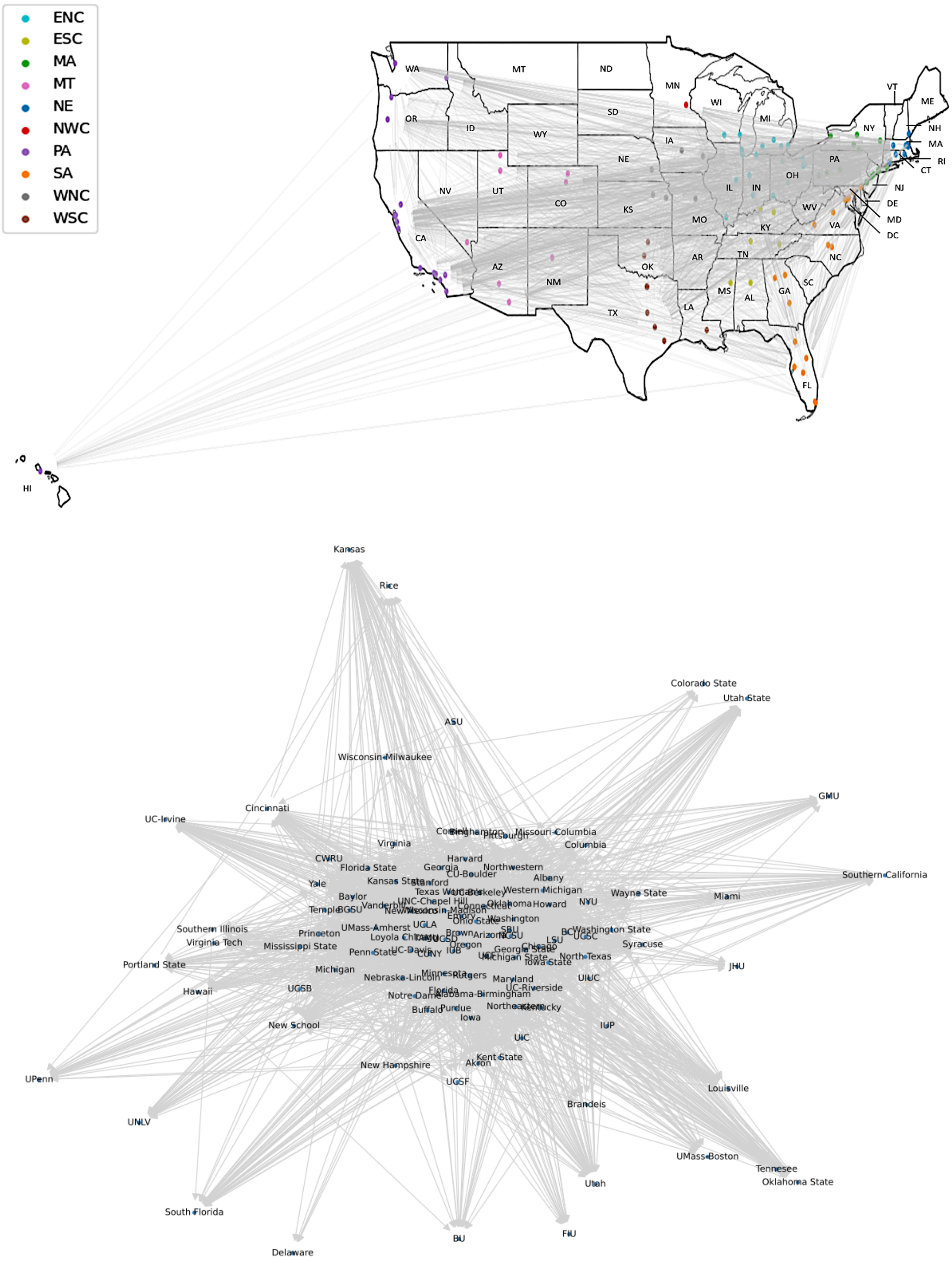

*Figure 4. Diffusion map across schools.*

## Knowledge diffusion in American sociology

Figure 4 above illustrates the geographical and network structure of research trends diffusion among sociology doctoral-granting institutions across the United States. The top panel presents a geographical map where nodes represent institutions that have either produced or adopted trendy terms, though not all schools are shown. Schools are color-coded by census region, highlighting regional clustering. The Northeast, Mid-West, and Western Coast regions exhibit the most substantial exchanges, indicating these areas' prominence in producing and adopting sociology research trends. Alaska is excluded as it has no sociology doctoral-granting institutions. The bottom panel showcases the directional and weighted network of vogue exchange among schools, revealing a clear core-periphery structure. Core institutions dominate the network, serving as the primary producers and adopters of vogue terms, with 77% of vogue flows occurring between core institutions. Peripheral institutions, identified using a strongly-connected-components detection algorithm (Hong et al., 2013) and comprising 17% of schools (e.g., 'Boston University,' 'Kansas,' 'Louisville,' 'George Mason'), play a limited role in this diffusion process. Specifically, 17% of flows move from core to peripheral institutions, 5% flow from peripheral to core, and only 1% occur between peripheral institutions.

Table 3 investigates the interaction effects between institutional classification, ranking, and geographical locations on the diffusion of research trends. The results from the six models in Table 3 yield several significant findings. Public schools located within the same U.S. Census sub-region are significantly more likely to adopt trendy terms produced by each other, with the interaction effect between public classification and the same location yielding a 0.088 increase (+8.77%) in adoption. In contrast, public schools that share the same prestige ranking exhibit a negative interaction effect, with same-ranking public school dyads obtaining a 0.134 decrease (-12.56%) in adoption, which may suggest competition among similarly ranked public institutions. Private schools, however, show a different dynamic. The interaction between private classification and the same ranking results in a 0.352 increase (+34.56%) in adoption, indicating that similarly ranked private schools are more inclined to exchange and adopt trendy terms.

Table 4 demonstrates that prior research fit, as measured by dissertation abstract text similarity, significantly predicts the adoption of research trends. The strong positive coefficient for fit across multiple models suggests that institutions are more likely to adopt research trends that align closely with their existing intellectual trajectories. The negative and significant coefficient for rank gap ($-0.005$, $p<0.01$) indicates that the greater the disparity in rank between the adopter and the producer, the less likely the adoption will occur.

| | *Dependent variable: The number of adopted pairs* | | | | | |
|---|---|---|---|---|---|---|
| | (1) | (2) | (3) | (4) | (5) | (6) |
| const | 0.146* | 0.070 | 0.069 | 0.127 | 0.064 | 0.065 |
| | (0.074) | (0.073) | (0.073) | (0.074) | (0.073) | (0.073) |
| both public | -0.105** | | | -0.083** | | |
| | (0.020) | | | (0.020) | | |
| both landgrant | | -0.009 | | | -0.004 | |
| | | (0.024) | | | (0.024) | |
| both private | | | 0.104** | | | 0.069 |
| | | | (0.037) | | | (0.036) |
| same location | -0.029 | 0.010 | 0.022 | | | |
| | (0.034) | (0.024) | (0.023) | | | |
| same rank | | | | 0.177** | 0.089** | 0.077** |
| | | | | (0.039) | (0.027) | (0.026) |
| location×public | 0.088* | | | | | |
| | (0.045) | | | | | |
| location×landgrant | | 0.068 | | | | |
| | | (0.063) | | | | |
| location×private | | | -0.038 | | | |
| | | | (0.086) | | | |
| rank×public | | | | -0.134** | | |
| | | | | (0.051) | | |
| rank×landgrant | | | | | 0.040 | |
| | | | | | (0.070) | |
| rank×private | | | | | | 0.352** |
| | | | | | | (0.112) |
| Observations | 11187 | 11187 | 11187 | 11187 | 11187 | 11187 |
| $R^2$ | 0.165 | 0.163 | 0.164 | 0.166 | 0.164 | 0.165 |
| Adjusted $R^2$ | 0.158 | 0.156 | 0.156 | 0.159 | 0.157 | 0.158 |
| Residual Std. Error | 0.771 | 0.772 | 0.772 | 0.771 | 0.772 | 0.771 |
| | (df=11085) | (df=11085) | (df=11085) | (df=11085) | (df=11085) | (df=11085) |
| F Statistic | 21.715** | 21.403** | 21.486** | 21.922** | 21.556** | 21.756** |
| | (df=101; 11085) | (df=101; 11085) | (df=101; 11085) | (df=101; 11085) | (df=101; 11085) | (df=101; 11085) |
| *Note:* | | | | | | *p<0.05; **p<0.01 |

***Table 3.*** *School-level Characteristics in Shaping Research Trends Diffusion*

| | *Dependent variable: The number of adopted pairs* | | | | | | |
|---|---|---|---|---|---|---|---|
| | (1) | (2) | (3) | (4) | (5) | (6) | (7) |
| const | -0.517** | -0.517** | -0.529** | -0.529** | -0.550** | -0.467** | -0.559** |
| | (0.076) | (0.076) | (0.093) | (0.093) | (0.094) | (0.078) | (0.096) |
| *fit* | 1.967** | 1.969** | 1.810** | 1.814** | 1.955** | 1.946** | 1.944** |
| | (0.072) | (0.072) | (0.115) | (0.115) | (0.149) | (0.072) | (0.149) |
| same location | | -0.013 | | -0.029 | -0.030 | | -0.029 |
| | | (0.022) | | (0.029) | (0.029) | | (0.029) |
| rank gap | | | -0.005** | -0.005** | -0.006** | | -0.006** |
| | | | (0.000) | (0.000) | (0.000) | | (0.000) |
| size gap | | | | | -0.001 | | -0.001 |
| | | | | | (0.001) | | (0.001) |
| both landgrant | | | | | | -0.032 | -0.064* |
| | | | | | | (0.024) | (0.032) |
| both private | | | | | | 0.043 | -0.012 |
| | | | | | | (0.035) | (0.041) |
| both public | | | | | | -0.058** | 0.013 |
| | | | | | | (0.021) | (0.027) |
| Observations | 10598 | 10598 | 7313 | 7313 | 7313 | 10598 | 7313 |
| $R^2$ | 0.221 | 0.221 | 0.263 | 0.264 | 0.264 | 0.222 | 0.264 |
| Adjusted $R^2$ | 0.214 | 0.214 | 0.255 | 0.255 | 0.255 | 0.214 | 0.255 |
| Residual Std. Error | 0.753 | 0.753 | 0.807 | 0.807 | 0.807 | 0.752 | 0.807 |
| | (df=10498) | (df=10497) | (df=7227) | (df=7226) | (df=7225) | (df=10495) | (df=7222) |
| F Statistic | 30.070** | 29.771** | 30.409** | 30.067** | 29.752** | 29.360** | 28.812** |
| | (df=99; 10498) | (df=100; 10497) | (df=85; 7227) | (df=86; 7226) | (df=87; 7225) | (df=102; 10495) | (df=90; 7222) |
| *Note:* | | | | | | | *p<0.05; **p<0.01 |

***Table 4:*** *Previous research fit in shaping Research Trends Diffusion*

Overall, our results support a model of differentiation rather than hierarchical diffusion. We find no evidence to show that lower status departments actively mimic the vogue produced by higher status departments in the sense of status conformity. Rather, our results suggest that the movement of knowledge production among the departments is more complicated than simply from the top-down hierarchical model. For instance, public institutions are more likely to distinguish themselves from each other, while private institutions are more likely to adopt each other's vogue. Elder & Kozlowski (2025) report that high-status sociology departments are more likely to invest in topics that have been frequently featured in high status journals, as it often is perceived as prestigious; while lower-status sociology departments focus more on applied topics that are closely related to current social debate. This adds further interpretation of our results, lower-ranked departments may be less likely to adopt trends originating from elite institutions, not solely due to institutional barriers but also because of a misalignment in research priorities. In this sense, diffusion occurs more readily among institutions with similar orientations, as they share methodological and thematic commitments. This helps explain why the adoption of research trends is not purely a hierarchical process but is shaped by the intellectual structure of the discipline itself. Adding upon Elder & Kozlowski, (2025) and Gondal (2018), we show that applied and social policy-driven topics, despite often being perceived as less prestigious than theoretical oriented topics, in fact leading the most vogue. This phenomenon is salient in both dissertation research and peer reviewed publications.

# Discussion

Bourdieu's *Homo Academicus* (1988) elaborates on the structuring forces within academia, where academic hierarchies are reproduced through networks of power, institutional affiliations, and access to research funding. In this framework, the dominance of high-status institutions in shaping research trends is not merely a reflection of merit but a product of institutionalized mechanisms that sustain the existing power structure. High-status institutions accumulate symbolic capital by generating and exchanging research trends primarily among themselves, much like high-status social actors use cultural tastes to differentiate themselves from lower-status groups. Lower-status institutions, in turn, seek to enhance their academic legitimacy by adopting trends initiated by elite institutions, reflecting the logic of social reproduction.

This study offers a contradictory explanation to this view, our results show that, rather than following a strict hierarchical model where lower-status institutions mimic elite research priorities, knowledge production in sociology follows a more segmented structure. Specifically, we find little overlap between the vogue terms of high- and low-status institutions, suggesting that lower-status schools do not simply adopt the intellectual agenda set by elite departments. Instead, these institutions cultivate distinct research niches that diverge from the trends dominating core institutions. Moreover, our analysis highlights the significant role of external factors, such as geographical location, in shaping knowledge production. Regional clustering and institutional classification appear to exert a stronger influence on intellectual trajectories than previously assumed, reinforcing the idea that knowledge production is not only structured by status hierarchies but also by contextual and spatial dynamics.

Our relational methodology captures research trends as living processes, shaped by the constant interplay of scholarly interactions, which challenges the traditional view that portrays the development of scientific ideas as a static process. By capturing how research themes rise, diffuse, and stabilize within sociology, this study provides a new perspective on the dynamic interplay of novelty, conformity, and adaptation in shaping intellectual trajectories, offering an alternative lens to complement and expand existing bibliometric approaches. Our findings further illustrate the role of meso-level institutional factors—geographical location, departmental prestige, and institutional classification—in shaping the pathways of intellectual exchange. These meso-level influences, often overlooked in prior research, reveal how conceptual and methodological shifts interact with meso-level institutional contexts to shape the diffusion of intellectual trends. For instance, public universities residing in the same geo-location tend to emphasize regionally salient issues, while no significant preference was found in private universities. These distinctions illuminate the complex interplay between institutional characteristics and the emergence of disciplinary research trends.

Leahey et al. (2023) analyze how different types of novelty in scientific work influence their capacity to disrupt established paradigms, identifying forms of innovation most likely to transform disciplinary knowledge using bibliometric measures like citation patterns. Park et al. (2023) demonstrate a trend where contemporary scientific papers and patents increasingly consolidate existing knowledge rather than displacing it, reflecting a shift in how research contributes to the evolution of scientific fields. This study examines the conceptual evolution of sociological research ideas and with a special focus on the semantic relations within texts to trace the movement and diffusion of intellectual trends. As outlined in Kedrick et al (2024), the production of scientific knowledge was operated through the hierarchical organization of concepts, where core concepts act as foundational anchors and peripheral concepts drive innovation. Core concepts are highly connected and stable, providing coherence across subfields, while peripheral concepts are less connected, dynamic, and introduce new ideas. This study provides empirical support for this relational view by mapping the core-peripheral structure of research concepts in American sociology. We demonstrate that research trends are more likely to be weakly connected within the semantic structure. Likewise, we show that topics on methodology and theory have the least overlap with the number of word-pairs that represent research trends. This aligns with Leahey et al. (2023), who show that novel methods and theories, while impactful when successful, are particularly difficult to establish due to their reliance on broad applicability and foundational shifts in thinking.

Elder and Kozlowski (2025) examine the relationship between departmental status and the distribution of subfields within American sociology. Their study reveals a clear stratification of subfields along institutional hierarchies, with elite departments disproportionately emphasizing male-dominated, theoretically oriented subfields such as *Theory* and *Economic Sociology*. In contrast, lower-status departments are more likely to focus on applied, female-dominated subfields such as *Health* and *Criminology*, which align with practical career pathways and societal applications. While Elder and Kozlowski attribute these patterns to a combination of gendered dynamics and the symbolic value of theoretical subfields, which confer prestige despite their limited practical relevance, our research offers a complementary perspective. We find that applied research, despite being perceived as low prestige, is actually the major driving force behind the emergence and diffusion of research trends within the discipline. Our analysis provides further evidence to illustrate how sociology Ph.D. granting institutions, based on their perceived departmental prestige, classification, and geo-location, develop different strategies for producing and adopting research trends in the discipline. Future research could explore how government funding policies, whether through support or restriction, influence the production and diffusion of knowledge within sociology and related disciplines. For instance, funding priorities that favor applied research addressing public health or criminal justice may further entrench the stratification between departments focusing on practical applications and those emphasizing theoretical subfields.

Our findings are subject to several limitations. First, our reliance on textual data—such as dissertation abstracts—may overlook other influential aspects of knowledge production, including unpublished work, informal discussions, and non-textual forms of intellectual engagement. The exclusion of informal networks, such as mentorship relationships and unrecorded collaborations, means that critical pathways of idea diffusion and adaptation remain unexplored. These informal interactions often play a significant role in shaping intellectual movements and deserve greater attention in future research. As a result, our analysis may present an incomplete picture of how ideas are developed and disseminated. Additionally, while our longitudinal approach is valuable for capturing trends over time, it may fail to account for rapid shifts in intellectual priorities or the long incubation periods that some concepts require before gaining prominence. This temporal constraint could limit our ability to fully trace the lifecycle of certain ideas. Moreover, because terms are largely shaped by the scholar-specific characteristics of graduates in a given year, there is considerable randomness in term usage from year to year. To measure intellectual vogues in the 2010s, it is natural to split the decade into two periods: the first half and the second half. Future research could extend our analysis over a longer time span to generate more robust and generalizable insights. More advanced techniques — such as hypergraph modeling (Shi and Evans 2023) — could also be employed to capture the dynamics of semantic networks.

This work makes a complementary contribution to the sociology of knowledge by extending existing models of academic institutional stratification and knowledge diffusion. Rather than viewing knowledge production as a process driven primarily by elite institutions through the top-down hierarchical mode, our analysis reveals a more complex and institutionally embedded dynamic. We further show that the production and circulation of research trends are not merely dictated by prestige but are shaped by the interplay of institutional classifications, geographical proximity, and department research specializations. This perspective aligns with broader theoretical debates on how intellectual authority is constituted and maintained, demonstrating that the landscape of knowledge production academia is continuously altered through internal scholarly activities and external socio-political environments.

**Acknowledgments**: We are grateful to the discussants at the International Conference for Computational Social Science 2024 and the International Conference on Science of Science and Innovation 2024. All errors are our own.

# Appendix

## 1. Foundations VS Trends in American Sociology

As described in the original paper we compare the vogue terms (moving from peripheral to backbone) and the foundational terms (always on the backbone across periods).

| Representative topics | Vogue | Foundation |
|---|---|---|
| 'health' | 'objective','distress','parental', 'planning','incarceration', 'stigma','cognitive','reproductive','socioeconomic','young','depressive' | 'age', 'later', 'medical', 'old', 'elder' |
| 'mental' | 'adult','physical','stress','black','psychosocial','discrimination','identity','distress','stigma', 'psychological','justice','adolescent','young' | mental has no connected edges on the first-period's backbone |
| 'gender' | 'religious','medical','violence','feminist','young','transgender','money' | 'sexual', 'women', 'sexuality', 'people', 'woman' |

A clear pattern emerges: vogue terms emphasize specific, timely social issues. Terms such as incarceration, reproductive, transgender, and psychosocial suggest a focused social context for these representative topics, whereas foundation terms reflect broader, less context-bound symbolic meanings.

## 2. What school-level characteristics make a school more likely to produce or adopt trending terms?

*(a) We examined the relationship between school-level characteristics and the production of vogue terms during the first five years of our study period. The dependent variable is the number of vogue terms produced by each institution. Ordinary Least Squares (OLS) regression was performed with institutional attributes including size, institutional classification, and average ranking as predictors. Institutions can be simultaneously classified as public and land-grant. The regression results show that institutional size is the only statistically significant predictor of vogue production ($\beta$ = 0.2495, $p < 0.001$). This indicates that larger institutions are more likely to produce vogue terms, possibly due to greater faculty numbers or research output capacity.*

| Variable | Coefficient | *P*-value | 95% CI |
|---|---|---|---|
| const | 1.2636 | 0.716 | [-5.626, 8.153] |
| size | 0.2495 | 0.000 | [0.201, 0.298] |
| private | -1.1844 | 0.733 | [-8.074, 5.705] |
| public | 2.083 | 0.539 | [-4.625, 8.791] |
| land_grant | -1.3425 | 0.109 | [-2.991, 0.306] |
| avg_ranking | -0.0095 | 0.558 | [-0.042, 0.023] |

*(b) This regression examines how institutional characteristics in the first five years predict the adoption of vogue terms in the subsequent five years. The dependent variable is the number of vogue terms adopted by each institution. OLS regression was performed with predictors including institutional size, classification (private/public/land-grant), and average prestige ranking. Results indicate that institutional size is positively and significantly associated with vogue adoption ($\beta$ = 0.1598, $p = 0.003$), suggesting that sociology departments with a larger PhD student size are more likely to adopt vogue. Meanwhile, the average ranking has a significant negative relationship with vogue adoption ($\beta$ = -0.1484, $p < 0.001$). Since a higher ranking score indicates a lower-status institution, this result shows that lower-ranked schools are less likely to adopt vogue terms, although these non-elite public schools produce them.*

| Variable | Coefficient | *P*-value | 95% CI |
|---|---|---|---|
| const | 27.4356 | 0.000 | [12.750, 42.122] |
| size | 0.1598 | 0.003 | [0.056, 0.264] |
| private | -12.6006 | 0.092 | [-27.288, 2.087] |
| public | -10.1046 | 0.164 | [-24.404, 4.195] |
| land_grant | -2.267 | 0.203 | [-5.782, 1.248] |
| avg_ranking | -0.1484 | 0.000 | [-0.217, -0.080] |